\documentclass[5p,twocolumn]{elsarticle}

\usepackage{comment}
\usepackage{amsmath,amsthm}
\usepackage{amssymb}
\usepackage{amsfonts}
\usepackage{mathtools}
\usepackage{mathrsfs}  
\usepackage{multirow}
\usepackage{algorithm}
\usepackage{algpseudocode}
\usepackage{dashrule}
\usepackage{booktabs}
\usepackage{adjustbox}
\usepackage{dcolumn}
\usepackage{cancel}

\usepackage{xurl}
\usepackage[hidelinks,breaklinks=true]{hyperref}

\usepackage{tikz}
\usetikzlibrary{shapes.geometric, arrows, positioning}

\usepackage[dvipsnames, table]{xcolor}

\usepackage{subfigure}

\usepackage{bm}

\usetikzlibrary{
    arrows.meta,
    backgrounds,
    fit,
    positioning,
    calc,
    shadows
}

\newcommand{\argmin}{\arg\min}

\def\R{{\mathbb{R}}}

\newcommand {\bsis} {\left\{ \begin{array} } 
\newcommand {\esis} {\end{array}\right.}

\def\bmat#1{\left[\begin{array}{#1}}
\def\emat{\end{array}\right]}

\newcommand{\blista}{\renewcommand{\labelenumi}{(\roman{enumi})} 
	\begin{enumerate}}
	\newcommand{\elista}{\end{enumerate} \renewcommand{\labelenumi}{\arabic{enumi}.}}

\newtheorem{thm}{Theorem}

\newtheorem{prop}[thm]{Proposition}

\tikzstyle{circleblock} = [
  circle,
  minimum size=1.5cm,
  text centered,
  draw=black
]
\tikzset{
  rhomboidblock/.style={
    trapezium,
    trapezium left angle=70,
    trapezium right angle=110,
    draw=black,
    minimum width=3.5cm,
    minimum height=1cm,
    text centered,
    align=center
  }
}

\tikzstyle{startstop} = [rectangle, rounded corners, minimum width=3cm, minimum height=1cm,text centered, draw=black]
\tikzstyle{process} = [rectangle, minimum width=3.5cm, minimum height=1cm, text centered, draw=black]
\tikzstyle{decision} = [diamond, aspect=2, minimum width=3cm, minimum height=1cm, text centered, draw=black]
\tikzstyle{arrow} = [thick,->,>=stealth]

\journal{Sustainable Energy, Grids and Networks}

\begin{document}

\begin{frontmatter}

\title{Accelerated kriging interpolation for real-time grid frequency forecasting\tnoteref{finalversion}}

\tnotetext[finalversion]{\copyright~2026 The Authors. This is an open access article under the CC BY license (\url{https://creativecommons.org/licenses/by/4.0/}). Final version of the article available at \url{https://doi.org/10.1016/j.segan.2026.102403} (please cite as \cite{moreno2026accelerated}).}

\author[a]{Carlos~Moreno-Blazquez\corref{cor1}}
\ead{cmb@us.es}
\author[a]{Filiberto~Fele}
\ead{ffele@us.es}
\author[a]{Teodoro~Alamo}
\ead{talamo@us.es}

\cortext[cor1]{Corresponding author}

\affiliation[a]{organization={Department of Systems Engineering and Automation, University of Seville},
             addressline={Av.~Camino de los Descubrimientos s/n},
             city={Seville},
             postcode={41092},
             country={Spain}}

\begin{abstract}
The integration of renewable energy sources and distributed generation in the power system calls for fast and reliable predictions of grid dynamics to achieve efficient control and ensure stability. In this work, we present a novel nonparametric data-driven prediction algorithm based on kriging interpolation, which exploits the problem's numerical structure to achieve the required computational efficiency for fast real-time forecasting. Our results enable accurate frequency predictions directly from measurements, achieving sub-second computation times. We validate our findings on a simulated distribution grid case study.
\end{abstract}


\end{frontmatter}

%============================================================================================================================================================================================================================================================================================================================
\section{Introduction}  
\label{sec:introduction}

The dynamic behavior of frequency encapsulates essential information about the energy balance in a power system. Rapid deviations---which may occur within seconds after a disturbance---reflect the complex interactions among various network elements \cite{MOHAMMADI2024QuantifyFrequency}. 
Maintaining the nominal frequency is essential for preventing cascading failures and costly blackouts \cite{yan2018anatomy, chen2020distributed}. 

The ongoing replacement of conventional synchronous generators by inverter-based resources (IBRs) leads to a reduction in system inertia, resulting in a faster rate of change of frequency (ROCOF) and new converter-driven stability dynamics \cite{hatziargyriou2020definition, CarEtAl2021_IEEETIE}.  

Advances in measurement devices---such as phasor measurement units (PMUs), which operate at sub-second resolutions and are synchronized via GPS \cite{Xu2015RealTimePMU, DOMINGUEZ2023EvolPowerSys}---have enabled operators to detect and analyze rapid frequency transitions. This sub-second visibility is particularly relevant given the operating speed of modern protection mechanisms. For instance, fast-response systems in North America activate in as little as 250 milliseconds \cite{cauley2020north, LAVI2023109422, Du2023}. Similarly, European grids are evolving beyond the traditional Frequency Containment Reserve by introducing Fast Frequency Response (FFR) mechanisms that operate in comparable sub-second timeframes \cite{eriksson2018synthetic} to maintain stability standards like EN50160 \cite{ss_en50160_2010}.

All this highlights the necessity for high temporal-resolution predictions that allow for early detection of imbalances and prompt activation of control measures, even over sub-second prediction horizons \cite{panagi2025sizing, milano2018foundations}.
Historically, models used to represent power-system dynamics have been based on physics-driven parametric formulations \cite{kundur2007power, yazdani2010voltage}.
However, these approaches often fail to capture the fast transients of modern grids. While model order reduction techniques attempt to mitigate the computational burden, they face significant challenges. For example, traditional linear methods compromise accuracy when renewable variability shifts the system's operating point \cite{milano2018foundations,muntwiler2024stiffness}. 
Conversely, more refined nonlinear models involve high computational costs for modeling and simulation \cite{kumar2023state}. Furthermore, most order reduction approaches necessitate high-fidelity physical models of both the network and the connected components, which are frequently unavailable due to proprietary manufacturer data \cite{milano2018foundations,nadeem2024structure}.

Research on frequency trajectory prediction and emergency control has mostly diverged into two main streams. Model-based methods often use techniques like trajectory sensitivity analysis but depend on accurate system parameters, which are often uncertain \cite{azizipanah2019modeling}. In contrast, purely data-driven approaches offer a pragmatic and robust alternative, as they learn the system dynamics directly from real-world measurements, thereby bypassing the need for an explicit physical model and inherently capturing the complex, nonlinear behaviors that characterize grid operation \cite{sharma2021data, bertozzi2024application, barocio2014dynamic, MarkovskyEtAl2023}.
A prominent body of work employs deep learning models to forecast the frequency trajectory, demonstrating high accuracy in capturing temporal dependencies \cite{chamorro2021data, CHETTIBI2021RealTimeFreq}. Other machine learning techniques, such as robust ensemble or white-box approaches, have been successfully applied to dynamic security assessment \cite{zhang2019online}, while solutions based on deep reinforcement learning are being developed for adaptive emergency control \cite{huang2019adaptive}.

While deep learning frameworks demonstrate powerful predictive capabilities, their application to safety-critical grid control is hindered by several fundamental limitations. First, their black-box nature prevents both principled uncertainty quantification and model diagnostics, which are essential for building trust with system operators. Second, deep learning models are notoriously data-hungry, requiring vast datasets that may not be available for the rare but critical contingency events that are of most interest, posing a risk of poor generalization when data is scarce. Finally, natively incorporating hard physical constraints is non-trivial in neural network architectures \cite{pineda2025beyond,de2025learning}.

These requirements motivate an evaluation of established interpolation and regression frameworks. However, many conventional interpolation techniques are themselves insufficient. Methods such as polynomial interpolation or deterministic splines impose rigid structural assumptions on the data, often failing to capture non-stationary behaviors. Other approaches like inverse distance weighting are purely geometric, assigning weights based on a fixed function of distance without considering the underlying correlation structure of the data itself \cite{li2014spatial}.

In contrast to these approaches, stochastic interpolation methods model the variable of interest as a realization of a random function defined by some covariance structure.
A well known method in this class is Gaussian process (GP) regression \cite{Rasmussen2006gaussian}, which has been largely considered in power systems literature for probabilistic stability analysis \cite{ye2022physics}, online dynamic security assessment \cite{zhai2022dynamic}, frequency dynamics inference \cite{jalali2022inferring}, and predictive control strategies \cite{gan2020data}. A distinct methodological emphasis is offered by kriging---a method rooted in geostatistics \cite{krige1981lognormal, matheron1967kriging}---which shares its theoretical foundation with GPs. However, while the latter stands on a Bayesian framework, which typically involves assumptions regarding data normality and second-order stationarity, kriging employs the so-called variogram to explicitly model the spatio-temporal dependence of the data: this relies on the intrinsic hypothesis---a less restrictive condition than second-order stationarity \cite{cressie2011statistics, matheron1963principles}. The output of kriging regression is an unbiased point prediction accompanied by a minimal prediction variance \cite{schulz2018tutorial}. Although this metric serves a different role than the full posterior distribution in Bayesian inference, it provides a functional measure of uncertainty to quantify the confidence in the forecast, which is valuable for risk-aware decision-making \cite{girard2002multiple}.

Despite its interesting properties and successful application in a dynamical systems context \cite{carnerero2021probabilistic, carnerero2023kernel}, the application of kriging for real-time frequency prediction remains a largely unexplored area. This is primarily due to numerical and theoretical hurdles that complicate its deployment. First, the method involves handling a dense covariance matrix that is often severely ill-conditioned due to high temporal correlations in the data \cite[A.~2]{Rasmussen2006gaussian}. Second, when constructed via the so-called variogram, this matrix is only \emph{conditionally positive definite}, meaning the problem is strictly convex only on some affine subspace, while it is indefinite elsewhere \cite[Sec.~2.3]{chiles2012geostatistics}. 
While standard kriging admits an analytical solution, it typically produces dense predictors susceptible to the screening effect (see \cite[Sec.~3.2.1]{cressie1993statistics} and \cite{stein2002screening})---where negative weights can be counter-intuitively assigned to redundant sensors. Introducing $\ell_1$-regularization resolves this by enforcing sparsity and facilitating interpretability, but it comes at a steep computational cost: it precludes the use of fast analytical solutions, requiring instead the solution of a non-differentiable optimization problem at every time step. 

This paper introduces a novel framework for robust, interpretable, and computationally efficient real-time frequency prediction. Our main contributions are threefold:
\begin{enumerate}
    \item We propose a \emph{regularized universal kriging framework} incorporating an $\ell_1$-norm penalty tailored for frequency trajectory forecasting. By enforcing sparsity on the kriging weights, this model effectively mitigates the screening effect inherent to standard kriging, ultimately favouring interpolation over extrapolation. Unlike black-box dense methods, our approach delivers a physically interpretable predictor along with a principled measure of uncertainty.
    
    \item We develop a specialized numerical solution algorithm based on the alternating direction method of multipliers (ADMM). We leverage spectral decomposition of the variogram matrix to diagonalize the quadratic term, enabling ADMM to split the complex optimization into computationally lightweight scalar operations. Moreover, the augmented Lagrangian structure of ADMM introduces a quadratic proximal penalty, which numerically stabilizes the possibly ill-conditioned Hessian matrix and ensures robust convergence.

    \item We demonstrate the feasibility of the proposed solver for real-time applications. The methodology transforms high-resolution grid measurements into accurate forecasts of future frequency trajectories within stringent operational time constraints, enabling the preemptive mitigation of frequency deviations.
\end{enumerate}

Our proposal builds on advances in nonparametric modeling \cite{alfonso2020stock, carnerero2023prediction} and state-of-the-art numerical optimization methods \cite{moreno2024prediccion}. The effectiveness of this approach is validated through a realistic case study simulating a non-stiff grid, characterized by reduced system inertia and high sensitivity to disturbances.

%============================================================================================================================================================================================================================================================================================================================
\section{Universal Kriging}
\label{sec:kriging} 
Given a set of observed data pairs $\mathcal{D} = \{(y_1, z_1), \ldots, (y_N, z_N)\}$ from a random process $y: \mathbb{R}^n \to \mathbb{R}$, where $y_i = y(z_i)$ for $i = 1,\ldots,N$, kriging predicts the value $y_0$ at a query point $z_0$ using a linear combination of the observed values:
\begin{equation} \label{eq:predictor} 
    \hat{y}_0 = \sum_{i=1}^{N} \lambda_i y_i.
\end{equation}
The kriging weights, $\lambda = (\lambda_i)_{i=1}^N\in\mathbb{R}^N$, are calculated to satisfy two criteria: the estimator must be \emph{unbiased}, and the variance of the prediction error must be \emph{minimal}; this is often referred to as best linear unbiased estimator (BLUE) \cite[Sec.~4.1.2]{cressie2011statistics}. While the kriging predictor can be mathematically equivalent to the mean prediction of a GP model under certain stationarity assumptions, its derivation as a BLUE does not require assuming a Gaussian distribution for the underlying random field, offering broader applicability for point estimation \cite{cressie1993statistics}.

%==============================================================================================================================================================
\subsection{The Universal Kriging Model and Unbiasedness Constraints}
\label{sec:UK_model}
This article focuses on universal kriging (UK), which is designed for processes where the mean, or \emph{trend}, $\mu(z)=\mathbb{E}[y(z)]$, is not constant but varies \emph{deterministically} as a function of the coordinates---typically spatial and/or temporal, but here representing the combined 
state and input of a dynamical system. While the trend can be modeled using complex functions (e.g., higher-degree polynomials), this work adopts a linear form.
This is a standard methodological choice, as it avoids the case where a trend model with too many degrees of freedom wrongly absorbs the correlation that should be addressed instead to the residuals \cite{chiles2012geostatistics, journel1989we}.
Thus, we take:
\begin{equation} \label{eq:trend_model}
    \mu(z) = \bar{c} + Cz,
\end{equation}
where $\bar{c}\in \R$ and $C \in \mathbb{R}^{1 \times n}$ are unknown coefficients. The random process is thus modeled as:
\begin{equation} 
    y(z) = \mu(z) + \delta(z) = \bar{c} + C z + \delta(z), \label{eq:model_2}
\end{equation}
where $\delta(z)$ is a zero-mean random process.

To ensure that the predictor $\hat{y}_0$ in \eqref{eq:predictor} is unbiased---i.e., $\mathbb{E}[y_0 - \hat{y}_0] = 0$---for any value of the unknown coefficients $\bar{c}$ and $C$, the weights $\lambda_i$ must satisfy a set of linear constraints, the so-called unbiasedness constraints.
The expected prediction error is:
\begin{align*}
    \mathbb{E}[y_0 - \hat{y}_0] & = \mathbb{E}\left[y_0 - \sum_{i=1}^N \lambda_i y(z_i)\right] \nonumber\\
    & = (\bar{c} + Cz_0) - \sum_{i=1}^N \lambda_i (\bar{c} + Cz_i) \nonumber\\
    & = \bar{c}\bigg(1 - \sum_{i=1}^N \lambda_i\bigg) + C\bigg(z_0 - \sum_{i=1}^N \lambda_i z_i\bigg).
\end{align*}

\noindent For this to be zero, the following constraints must hold:
\begin{equation} \label{eq:uk_constraints}
    \sum_{i=1}^{N} \lambda_i = 1, \quad\sum_{i=1}^{N} \lambda_i z_i = z_0.
\end{equation}

%==============================================================================================================================================================
\subsection{Expressing Error Variance via the Semivariogram}
Under the unbiasedness constraints, we aim to minimize the variance of the prediction error $e \coloneqq y_0 - \hat{y}_0$, i.e., 
\begin{displaymath}
\operatorname{Var}[e] \coloneqq \mathbb{E}\left[(e-\mathbb{E}[e])^2\right]=\mathbb{E}\left[e^2\right],  
\end{displaymath}
where the latter equality follows from unbiasedness of the prediction, thus recovering the expression of the mean squared prediction error.
Let $\delta_i \coloneqq \delta(z_i)$, $i=0, \dots, N$. Using \eqref{eq:model_2} and \eqref{eq:uk_constraints}, we can rewrite the error as:
\begin{align} 
    e &= y_0 - \hat{y}_0 = (\bar{c} + C z_0 + \delta_0) - \sum_{i=1}^N \lambda_i (\bar{c} + C z_i + \delta_i) \nonumber \\
      &= \bar{c}\cancel{\bigg(1 - \sum_{i=1}^N \lambda_i\bigg)} + C\cancel{\bigg(z_0 - \sum_{i=1}^N \lambda_i z_i\bigg)} + \delta_0 - \sum_{i=1}^N \lambda_i \delta_i, \nonumber
\end{align}
from which we obtain
\begin{align} \label{eq:e^Te_expanded}
    \mathbb{E}[e^2] &= \mathbb{E}\left[\bigg(  \delta_0 - \sum_{i=1}^N \lambda_i \delta_i \bigg)^2\right] = \mathbb{E}\left[\bigg( \sum_{i=1}^{N} \lambda_i (\delta_0 - \delta_i) \bigg)^2\right] \nonumber \\
    &= \mathbb{E}\left[\bigg( \sum_{i=1}^{N} \lambda_i (\delta_0 - \delta_i) \bigg) \bigg( \sum_{i=j}^{N} \lambda_j (\delta_0 - \delta_j) \bigg)\right] \nonumber\\
    &= \mathbb{E}\left[\sum_{i=1}^{N}\sum_{i=j}^{N} \lambda_i\lambda_j  (\delta_0 - \delta_i)(\delta_0 - \delta_j)\right]\nonumber\\
    &= \sum_{i=1}^{N} \lambda_i \mathbb{E}\left[(\delta_0 - \delta_i)^2\right]  - \frac{1}{2} \sum_{i=1}^{N} \sum_{j=1}^{N} \lambda_i \lambda_j \mathbb{E}\left[ (\delta_i - \delta_j)^2\right],
\end{align} 
where the last equality is obtained by exploiting \eqref{eq:uk_constraints} and the linearity of $\mathbb{E}[\cdot]$, and reorganizing the terms.

A common approach to characterize $\mathbb{E} \left[ (\delta_i - \delta_j)^2\right] = \operatorname{Var} \left[\delta_i - \delta_j\right]$ with kriging is by relying on the \emph{intrinsic hypothesis}. Given any two coordinates $z,z'\in\mathbb{R}^n$, this hypothesis only assumes that $\operatorname{Var}[\delta(z')-\delta(z)]$ (hence, by unbiasedness, the variance of $y(z')-y(z)$) is finite and depends solely on the distance $h=z'-z$. This results in the standard definition of the \emph{semivariogram} $\gamma(h)$:
\begin{equation}\label{eq:th_variogram}
\gamma(h) \coloneqq \frac{1}{2} \operatorname{Var} \left[ \delta(z+h) - \delta(z) \right].
\end{equation}

Then, using \eqref{eq:th_variogram} in \eqref{eq:e^Te_expanded}, the variance of the prediction error is expressed as: 
\begin{equation} \label{eq:mspe_variogram}
\mathbb{E}[e^2] = 2 \sum_{i=1}^{N} \lambda_i \gamma(z_0 - z_i) - \sum_{i=1}^{N} \sum_{j=1}^{N} \lambda_i \lambda_j \gamma(z_i - z_j).
\end{equation}

We note that the semivariogram can be related to the covariance function, $C(h)$, used in models that assume second-order stationarity (as typical with GPs). Indeed, for such processes, it holds $\gamma(h) = C(0) - C(h)$.
Nonetheless, the semivariogram remains well defined even when $C(0)$ is infinite---an issue that arises when the global variance diverges due to a nonstationary mean \cite[Sec.~4.1.1]{cressie2011statistics}. We exemplify the empirical characterization of the semivariogram in Sec.~\ref{sec:variogram_model_cs}.

%==============================================================================================================================================================
\subsection{Universal Kriging Formulation}
\label{sec:UK_formulation}
The optimal kriging weights are those that minimize \eqref{eq:mspe_variogram}. Let us define the vector $\Gamma_0 \in \R^N$ and the symmetric matrix $\Gamma_\mathcal{D} \in \R^{N \times N}$ from the semivariogram values:
\begin{displaymath}
(\Gamma_0)_i = \gamma(z_0 - z_i), \quad (\Gamma_\mathcal{D})_{ij} = \gamma(z_i - z_j).
\end{displaymath}
Then, \eqref{eq:mspe_variogram} can be rewritten as:
\begin{displaymath}
\mathbb{E}[e^2] = 2\Gamma_0^\top \lambda - \lambda^\top \Gamma_\mathcal{D} \lambda,
\end{displaymath}
leading to the following equality-constrained quadratic program (QP):
\begin{subequations}\label{eq:OP_standard_QP}
    \begin{align}
    \min_{\lambda \in \mathbb{R}^N} 
    \ & -\lambda^\top \Gamma_\mathcal{D} \lambda + 2\Gamma_0^\top \lambda
    \label{eq:OP_standard_QP_a} \\[4pt]
    \text{subject to} \
    & R\lambda = r_0, \label{eq:OP_standard_QP_b}
    \end{align}
\end{subequations}
where $r_0 = [z_0^\top \; 1]^\top$, $r_i = [z_i^\top \; 1]^\top$, $R = [r_1, \ldots, r_N]$.
Provided an admissible variogram model is used \cite{christakos1984problem}, 
matrix $-\Gamma_\mathcal{D}$ is \emph{conditionally positive semidefinite} with respect to the unbiasedness constraints. This property is less strict than requiring $-\Gamma_\mathcal{D}$ to be positive semidefinite, as it implies that the objective function is convex only over the feasible set defined by \eqref{eq:OP_standard_QP_b} \cite[Sec.~2.3.2]{cressie1993statistics}.
Moreover, if the data points in $\mathcal{D}$ are sampled from distinct coordinates, $-\Gamma_\mathcal{D}$ is conditionally positive \emph{definite} and \eqref{eq:OP_standard_QP} has a unique minimum $\lambda^*$.

%==============================================================================================================================================================
\subsection{Regularized Universal Kriging}
To enhance the predictor's statistical generalization capabilities, we incorporate an $\ell_1$ regularization penalty in \eqref{eq:OP_standard_QP} which promotes sparsity and robustness \cite{HastieEtAl_Lasso2015Book}. Then, the optimal weights $\{\lambda_i^*\}_{i=1}^N$ are found by solving:
\begin{equation}\label{eq:OP_regularized}
    \begin{aligned}
    \lambda^* = \arg\min_{\lambda \in \R^N} \ & -\lambda^\top \Gamma_\mathcal{D} \lambda + 2\Gamma_0^\top \lambda + \sum_{i=1}^N \beta_i |\lambda_i| \\
    \text{subject to} \ & R\lambda = r_0,
    \end{aligned}
\end{equation}
where $\beta = (\beta_i)_{i=1}^N$, $\beta_i \ge 0$, controls the strength of the regularization. Interestingly, the $\ell_1$ penalty term interacts with the unbiasedness constraint $\sum \lambda_i=1$ to encourage non-negative solutions \cite{carnerero2023prediction, matsui2025sparse}. This mitigates the magnitude of negative weights, yielding a physically consistent model dominated by positive contributions from significant data points. Consequently, the predictor is biased towards \emph{interpolation rather than extrapolation}: this is well justified in the context of kriging, since the use of localized data aligns better with the hypothesis of stationarity of the mean in \eqref{eq:model_2}; see also \cite[Chapter~2]{multpointgeostatBOOK2014} and \cite{matsui2025sparse}.

As it will be shown through numerical experiments (Sec.~\ref{sec:Case_study}), this reduction in model complexity does not compromise accuracy; the proposed sparse estimation achieves an accuracy comparable to the non-regularized kriging approach.

%============================================================================================================================================================================================================================================================================================================================
\section{Efficient Numerical Solution}
While the regularized UK problem in \eqref{eq:OP_regularized} is convex whenever an admissible variogram is used (see Sec.~\ref{sec:UK_formulation}), the dense matrix $\Gamma_\mathcal{D}$ in the quadratic term and the non-differentiable $\ell_1$-norm render its solution computationally challenging for large datasets. To develop a numerically efficient scheme, we first reformulate the problem into a more tractable structure.

\subsection{Spectral Decomposition}\label{sec:spectral_decomp}
The main ingredient of our approach is the spectral decomposition of the Hessian matrix, $-\Gamma_\mathcal{D}$. Since the latter is real and symmetric, the spectral theorem ensures the existence of the decomposition $-\Gamma_\mathcal{D} = Q D Q^\top$, where $D = \mathrm{diag}(d_1,\dots,d_N)$ contains the eigenvalues, and the orthogonal matrix $Q\in\R^{N\times N}$ the associated eigenvectors. It is worth emphasizing that, for any semivariogram matrix, conditional positive definiteness implies the existence of a negative eigenvalue.

By introducing the change of variable $\nu = Q^\top\lambda$ into \eqref{eq:OP_regularized}, 
we obtain the equivalent diagonalized problem:
\begin{subequations}\label{eq:OP}
\begin{align}
   \min_{\nu\in\R^N} \ & \nu^\top D\nu  + 2\,\xi^\top\nu
    + \sum_{i=1}^N \beta_i |(Q\nu)_i|, \label{eq:diag-obj} \\
  \text{subject to} \
    & \tilde{R}\nu = r_0, \label{eq:OPcompact2}
\end{align}
\end{subequations}
where $\xi = Q^\top\Gamma_0$ and $\tilde{R} = R Q$.
This transformation offers a significant computational advantage. The quadratic term, $\nu^\top D\nu = \sum_{i=1}^N d_i \nu_i^2$, is now separable, although the decision variables are still coupled through the dense matrix $Q$ in the regularization term.

%==============================================================================================================================================================
\subsection{Numerical Challenges}
\label{sec:Challenges_and_ADMM_Case}
While the reformulation \eqref{eq:OP} offers a more structured problem, its numerical solution is far from trivial.
The eigenvalues of $-{\Gamma}_\mathcal{D}$ (contained in the diagonal matrix $D$), often reveal severe ill-conditioning: they can span many orders of magnitude, with many values being very close to zero. This reflects the high correlation between nearby data points and is a well-documented issue in kernel-based methods \cite{Rasmussen2006gaussian, schaback1995error}. This makes the problem sensitive to small perturbations and can lead to unstable solutions.

This challenge---together with the circumscribed convexity caused by conditional positive definiteness of the Hessian---poses a significant obstacle for standard, off-the-shelf optimization solvers. General-purpose quadratic programming solvers often assume global convexity and can struggle when the Hessian is indefinite outside the feasible set \cite{nocedal2006numerical}. Numerical inaccuracies can push the iterative solution outside the feasible hyperplane, potentially leading to physically meaningless solutions \cite{gould2002numerical}. Finally, the presence of the non-differentiable $\ell_1$-norm further complicates the use of gradient-based methods.

In the following section, we provide a detailed derivation of our solution to these challenges, building upon an ADMM scheme.

%============================================================================================================================================================================================================================================================================================================================
\section{K-ADMM Formulation}
\label{sec:ADMM_Formulation}
By introducing the auxiliary variable ${\alpha}\in\mathbb{R}^N$, \eqref{eq:OP} can be equivalently cast as
\begin{subequations}\label{eq:OP_ADMM}
\begin{align}
\min_{\nu, {\alpha}} & \quad \nu^\top D \nu - 2 \xi^\top \nu + \sum_{i=1}^N \beta_i |{\alpha}_i| \label{eq:OP_ADMM_cost}\\
\text{s.t.} & \quad \tilde{R} \nu = r_0, \label{eq:OP_ADMM_a}\\
& \quad Q \nu = {\alpha}, \label{eq:OP_ADMM_b}
\end{align}
\end{subequations}
obtaining a structure well suited for ADMM \cite{boyd2011distributed}. Indeed, \eqref{eq:OP_ADMM} is a minimization of $f(\nu) + g({\alpha})$ subject to the linear constraint \eqref{eq:OP_ADMM_b}, where:\footnote{We denote by $\mathcal{I}(\cdot)$ the indicator function, which returns zero if the condition expressed by the argument is met and $+\infty$ otherwise.}
\begin{align*}
    f(\nu) &= \nu^\top D \nu - 2 \xi^\top \nu + \mathcal{I}(\tilde{R} \nu = r_0), \\
    g({\alpha}) &= \sum_{i=1}^N \beta_i |{\alpha}_i|.
\end{align*}

From this, we obtain the augmented Lagrangian $\mathscr{L}_{\rho} : \mathbb{R}^{N} \times \mathbb{R}^{N} \times \mathbb{R}^{N} \rightarrow \mathbb{R}$,
\begin{multline}
    \mathscr{L}_{\rho}(\nu, {\alpha}, \eta) = \nu^\top D \nu - 2 \xi^\top \nu + \sum_{i=1}^N \beta_i |{\alpha}_i| + \mathcal{I}(\tilde{R} \nu = r_0)\\
    + \eta^\top (Q\nu - {\alpha}) + \frac{\rho}{2} \|Q\nu - {\alpha}\|_2^2,
    \label{eq:AugmentedLagrangian}
\end{multline}
which is parameterized by $\rho > 0$, and where $\eta \in \mathbb{R}^N$ is the vector of Lagrange multipliers associated with \eqref{eq:OP_ADMM_b}.
Given initial values ${\alpha}^0$ and $\eta^0$, the ADMM algorithm consists of the following sequence of iterations \cite[Sec.~3.1]{boyd2011distributed}:
\begin{subequations}\label{eq:ADMM_steps}
\begin{align}
    \nu^{k+1} &:= \argmin_{\nu \in \mathbb{R}^N} \ \mathscr{L}_{\rho}(\nu, {\alpha}^k, \eta^k), \label{step:v:plus}\\
    {\alpha}^{k+1} &:= \argmin_{{\alpha} \in \mathbb{R}^N} \ \mathscr{L}_{\rho}(\nu^{k+1}, {\alpha}, \eta^k), \label{step:lambda:plus}\\
    \eta^{k+1} &:= \eta^k + \rho (Q\nu^{k+1} - {\alpha}^{k+1}). \label{step:eta:plus}
\end{align}
\end{subequations}
Notice how the quadratic, differentiable part of problem \eqref{eq:OP_ADMM} is now decoupled from the non-differentiable and non-separable term, into \eqref{step:v:plus} and \eqref{step:lambda:plus}, respectively. We next detail the solution to each of these subproblems.

\subsection{The $\nu$-Subproblem}
The update for $\nu$ in \eqref{step:v:plus} involves minimizing the terms in \eqref{eq:AugmentedLagrangian} that depend on $\nu$, while keeping the hard constraint defined by \eqref{eq:OP_ADMM_a}. The optimization problem is the QP
\begin{align*}
\nu^{k+1} = \argmin_{\nu} \ & \nu^\top D \nu - 2 \xi^\top \nu + (\eta^k)^\top Q\nu + \frac{\rho}{2} \|Q\nu - {\alpha}^k\|_2^2 \\
\text{s.t.} \ & \tilde{R} \nu = r_0,
\end{align*}
whose solution must satisfy the KKT optimality conditions: 
\begin{equation}
\begin{bmatrix}
    2D + \rho Q^\top Q & \tilde{R}^\top \\
    \tilde{R} & \mathbf{0}
\end{bmatrix}
\begin{bmatrix}
    \nu^{k+1} \\
    \varsigma
\end{bmatrix}
=
\begin{bmatrix}
    2\xi - Q^\top\eta^k + \rho Q^\top{\alpha}^k \\
    r_0
\end{bmatrix},
\label{eq:KKT_system_v}
\end{equation}
where $\varsigma \in \mathbb{R}^{n+1}$ is the dual variable.
The top-left block of the KKT matrix (the matrix on the left-hand side) reveals the regularization introduced by the proximal penalty term in \eqref{eq:AugmentedLagrangian} (the term simplifies to $2D + \rho \mathbf{I}$ due to orthogonality of $Q$).
This significantly improves the conditioning of the $\nu$-subproblem's Hessian ensuring that each step of the algorithm is numerically stable, even when the original problem is nearly singular. Furthermore, it is worth noting that the KKT matrix is constant across iterations.
Then, $\nu^{k+1}$ can be retrieved efficiently by leveraging its pre-computed LU factorization, requiring at each iteration only forward-backward substitutions.

%==============================================================================================================================================================
\subsection{The {${\alpha}$}-Subproblem}
Problem \eqref{step:lambda:plus} reads as: 
\begin{equation*}
{\alpha}^{k+1} = \argmin_{{\alpha} \in \mathbb{R}^N} \ \sum_{i=1}^N \beta_i |{\alpha}_i| - (\eta^k)^\top {\alpha} + \frac{\rho}{2} \|Q\nu^{k+1} - {\alpha}\|_2^2.
\end{equation*}
This problem is separable.
By expanding the quadratic term, each component ${\alpha}_i$ is obtained as:
\begin{equation}
{{\alpha}_i^{k+1} = \argmin_{{\alpha}_i \in \mathbb{R}} \ \frac{\rho}{2} {\alpha}_i^2 + \beta_i |{\alpha}_i| - \left(\rho(q_i^\top \nu^{k+1}) + \eta_i^k \right) {\alpha}_i,}
\label{eq:lambda_separable}
\end{equation}
where $q_i^\top$ is the $i$-th row of $Q$. We note that \eqref{eq:lambda_separable} admits a solution in closed form, as detailed next; a proof is provided in \ref{Appendix:Proof:Optimal:z}.
\begin{prop}\label{prop:optimal:z}
    Consider the optimization problem 
    \begin{equation} \label{eq:family_problem}
         {x^* = \argmin_{x \in \mathbb{R}} \ w x^2 + \beta|x| - cx,}
    \end{equation}
    with $w>0$, $\beta\geq 0$, and $c\in \mathbb{R}$. Then, 
    \begin{equation} \label{eq:explicit_solution}
        {x^* = \psi(w,\beta,c) \coloneqq \frac{1}{2w} \mathrm{sign}(c) \max(0, |c| - \beta).}
    \end{equation}
\end{prop}
By comparing \eqref{eq:lambda_separable} with the general form in \eqref{eq:family_problem}, we directly identify the parameters specific to our problem:
\begin{equation} \label{eq:paremeters_prop}
w = \frac{\rho}{2}, \quad \text{and} \quad c_i = \rho(q_i^\top \nu^{k+1}) + \eta_i^k.
\end{equation}
Therefore, ${\alpha}_i^{k+1}$ can be computed directly through \eqref{eq:explicit_solution}; 
this step is computationally very inexpensive.

%==============================================================================================================================================================
\subsection{The Dual Update}
The update for the dual variable $\eta$ in \eqref{step:eta:plus} is a simple gradient ascent step, which ensures that the primal residual $\|Q\nu - {\alpha}\|$ converges to zero.
The complete iterative procedure of the proposed algorithm---hereafter referred to as K-ADMM---is summarized in Alg.~\ref{alg:ADMM_new}. The stopping criterion requires the primal and dual residuals to fall below predefined tolerances $\epsilon^{\text{pri}}$ and $\epsilon^{\text{dual}}$ \cite[Sec.~3.3.1]{boyd2011distributed}.

We remark that the penalty parameter $\rho$ governs the algorithm's convergence rate besides controlling the numerical conditioning of \eqref{eq:KKT_system_v}. Its selection remains an open problem in the literature: in practice, heuristic or adaptive strategies are commonly adopted. A detailed discussion on parameter tuning is beyond the scope of this work; we refer the reader to \cite[Sec.~3.4]{boyd2011distributed} for details.

%==============================================================================================================================================================
\subsection{Recovering the Solution}\label{sec:ADMM_out}
Upon convergence, Alg.~\ref{alg:ADMM_new} yields both $\nu^*$ and the auxiliary variable ${\alpha}^*$. While the splitting constraint \eqref{eq:OP_ADMM_b} theoretically implies that ${\alpha}^*$ corresponds to the kriging weights $\lambda^*$ (see  Section~\ref{sec:spectral_decomp}), in practice---due to the finite numerical tolerance $\epsilon^{\text{pri}}$---
${\alpha}^*$ complies with the unbiasedness constraints \eqref{eq:OP_ADMM_a} only approximately. In contrast, the update of $\nu$ involves solving the linear system \eqref{eq:KKT_system_v} which
ensures \eqref{eq:OP_ADMM_a} is met to machine accuracy.

Therefore, to prioritize the statistical validity of the estimator, we recover the final weights via the inverse transformation $\lambda^* = Q\nu^*$. If strict sparsity is required, a final post-processing step can be applied to truncate values of $\lambda^*$ below a negligible threshold to zero.

\begin{algorithm}[t]
    \caption{K-ADMM for Problem~\eqref{eq:OP_ADMM}} \label{alg:ADMM_new}
    \begin{algorithmic}[1]
    \State \textbf{Inputs:} Problem data $D, \xi, \beta, \tilde{R}, r_0, Q$. Parameters $\rho > 0, \epsilon^{\text{pri}} > 0, \epsilon^{\text{dual}} > 0$.
    \State \textbf{Initialization:} ${\alpha}^0 = \mathbf{0}$, $\eta^0 = \mathbf{0}$, $k = 0$.
    \Repeat
        \State Solve for $\nu^{k+1}$ by solving the linear system in \eqref{eq:KKT_system_v} using the pre-computed LU factorization.
        \State {Update ${\alpha}^{k+1}$ by applying \eqref{eq:explicit_solution} element-wise:
            \Statex \quad \,\, ${\alpha}^{k+1}_i \leftarrow \psi\big(\rho/2, \beta_i, (\rho Q\nu^{k+1} + \eta^{k})_i\big), \,\, \forall i=1,\dots,N$.
            }
        \State Update the dual variable $\eta^{k+1}$:
            \Statex \qquad\qquad $\eta^{k+1} \leftarrow \eta^{k} + \rho (Q\nu^{k+1} - {\alpha}^{k+1})$.
        \State Calculate $r^{k+1} \leftarrow \|Q\nu^{k+1} - {\alpha}^{k+1}\|_2$.
        \State Calculate $s^{k+1} \leftarrow \|\rho Q^\top({\alpha}^{k+1} - {\alpha}^{k})\|_2$.
        \State $k \leftarrow k + 1$.
    \Until{$r^k \leq \epsilon^{\text{pri}}$ and $s^k \leq \epsilon^{\text{dual}}$}
    \State \textbf{Output:} $\nu^* \leftarrow \nu^k$, ${\alpha}^* \leftarrow {\alpha}^k$.
    \end{algorithmic}
\end{algorithm}

\begin{figure*}[tb]
    \centering 
	\includegraphics[width=0.97\textwidth]{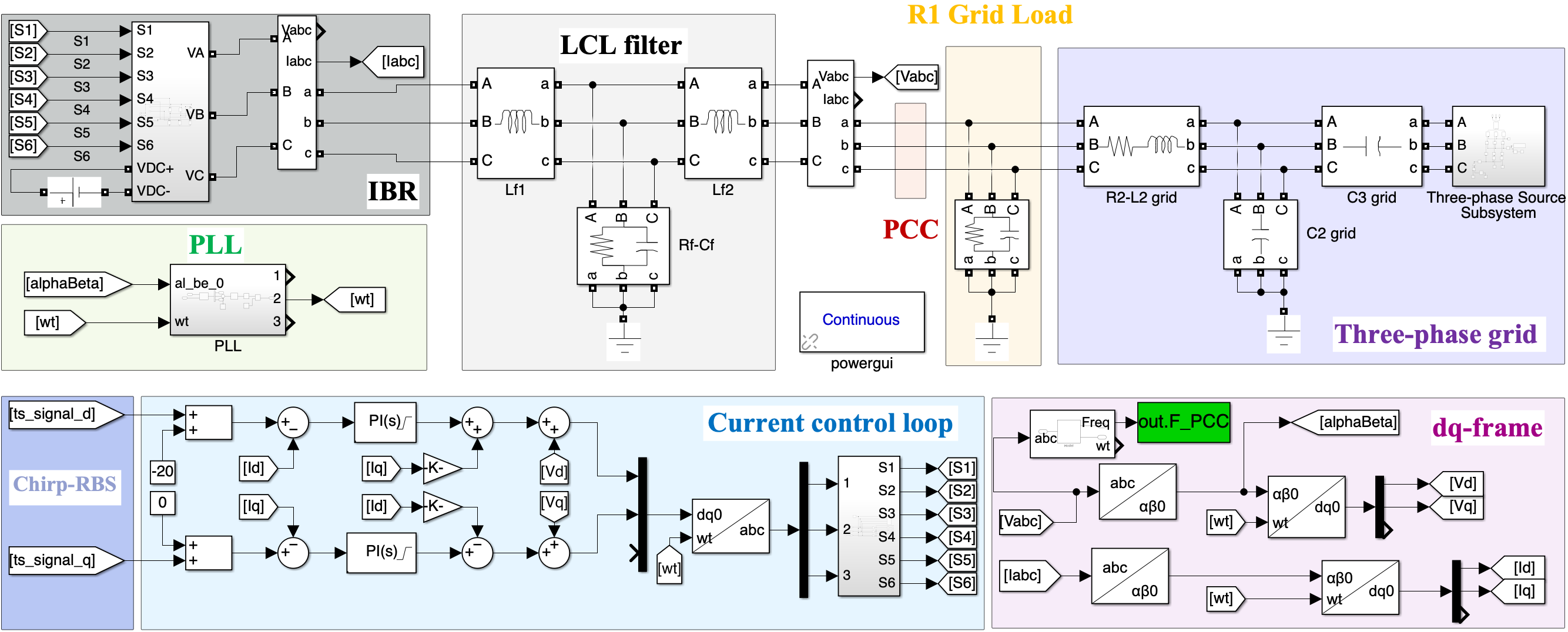}
        \caption{MATLAB's Simscape model of the proposed case study; the setup is built on the one proposed in \cite{haberle2023mimo}. Chirp-PRS signals were used as reference for the IBR to generate the dataset for the kriging-based prediction. }
	\label{fig:simulink}
\end{figure*}

%============================================================================================================================================================================================================================================================================================================================
\section{Case Study}
\label{sec:Case_study}
We evaluated the proposed kriging approach in a simplified setup which models power injections from an IBR at a point of common coupling (PCC)---where the distribution network interfaces with the main grid. The results demonstrate the effectiveness of our approach in capturing transient dynamics at sub-second scales, predicting frequency trajectories over a time horizon of $T = 0.5$ seconds. The latter is consistent with the $250$--$500$ ms operating range of FFR mechanisms \cite{eriksson2018synthetic}, providing the necessary buffer to compensate for sensing, computation and communication latencies \cite{milano2018foundations}; at the end of this section, we show how the computation times associated with Alg.~\ref{alg:ADMM_new} are compatible with such real-time applications.

\begin{table}[t]
    \centering
    \begin{footnotesize}
    \begin{tabular}{l c c}
    \toprule
    \textbf{Parameter} & \textbf{Symbol} & \textbf{Value} \\ \midrule
    Base Values & $V_b, S_b, f_b$ & 400 V, 1.5 kVA, 50 Hz \\ \hline
    Filter Values & $L_{f1}, L_{f2}$ & 0.027 p.u., 0.008 p.u. \\ \hline
    Filter Values & $C_f, R_f$ & 0.727 p.u., 0.031 p.u. \\ \hline
    Load Resistance & $R_1, C_1$ & 2 p.u., 0.05 p.u. \\ \hline
    Line & $R_2, L_2, C_2$ & 0.015 p.u., 0.15 p.u., 0.05 p.u. \\ \hline
    Main grid & $R_3, L_3, C_3$ & 0.015 p.u., 0.20 p.u., 10 p.u. \\ \hline
    Sampling Freq. & $f_s$ & 80 Hz \\ \hline
    Predict. Horizon & $n_p$ & 40 (500 ms) \\ \bottomrule
    \end{tabular}
    \end{footnotesize}
    \caption{Electrical parameters of the numerical experiment.}
    \label{tab:electrical_parameters}
\end{table}

The simulation was implemented in MATLAB's Simscape (see Fig.~\ref{fig:simulink}). We considered a weak grid to reflect the sensitivity of frequency and voltage to load variations, driven by a non-ideal three-phase source perturbed by stochastic noise, tuned to emulate realistic $\pm 20$~mHz frequency fluctuations.\footnote{The simulated grid incorporates an inductive transmission line ($R_2, L_2$) with shunt capacitance $C_2$, coupled with a non-ideal source ($R_3, L_3$) through the series compensation capacitance $C_3$. This yields a short-circuit ratio of $3.97$, which classifies it as a weak grid~\cite{de2019criterios}.} %
A proportional-integral control loop with current feedback was used to control the inverter, considering possible saturation of the actuators. The sampling rate was set to $f_s = 80\ \text{Hz}$, a decade above the cut-off frequency of the system; this was estimated by means of FFT analysis. The electrical parameters used in this case study are listed in Table~\ref{tab:electrical_parameters}.\footnote{A complete description of the electrical circuit, together with the MATLAB sources and the experimental dataset, is available in the Supplementary Material accompanying the final published version of this article \cite{moreno2026accelerated}; see Appendix B therein.} With this sampling rate, the $500$-ms time horizon corresponds to $n_p = 40$ discrete steps.

The dataset comprises historical sensor data---specifically, current injections in the direct-quadrature (\emph{dq}) reference frame, and frequency. We considered an ARX description of the system dynamics (autoregressive model with exogenous inputs), i.e.,
\begin{equation}\label{eq:ARX}
    y(t+1) =  F\bigl(y(t), \ldots, y(t - n_a),\,u(t), \ldots, u(t - n_b)\bigr) + e(t),
\end{equation}
where $y(t)\in\mathbb{R}$ is the grid frequency in Hz at time instant $t$, the vector $u(t)\coloneqq [i_d(t),\,i_q(t)]^\top \in \mathbb{R}^2$ represents the current inputs in the \emph{dq}-frame, and $e(t)$ accounts for modeling errors and process noise. Following a non-parametric approach, we postulate that \eqref{eq:ARX} holds for some unknown function $F$. The only specified parameters are the model orders, $n_a, n_b \in\mathbb{N}$, which define the number of past output and input values (regressors) used for prediction.

We generate the $n_p$-step ahead frequency trajectory by recursively applying a one-step-ahead predictor of the form 
\begin{equation}\label{eq:pred_ARX}
    \hat{y}(t+l+1\mid t) = F\bigl(z(t+l\mid t)\bigr),
\end{equation}
for $0\leq l< n_p-1$ \cite{ZHANG2004309}.
In \eqref{eq:pred_ARX}, the regressor vector $z(t+l\mid t)$ collects the past output and input values at prediction step $l$, as
\begin{multline}
    z(t+l\mid t) \coloneqq \bigl[ y(t+l\mid t), \ldots, y(t+l - n_a\mid t), \\
    u(t+l\mid t), \ldots, u(t+l - n_b\mid t) \bigr]^\top, \nonumber
\end{multline}%
where, for $0\leq m\leq n_a$,
\begin{displaymath}
    y(t+l-m\mid t)\coloneqq
    \begin{cases}
        \hat{y}(t+l-m\mid t), & \text{if } t+l-m > t,\\
        y(t+l-m), & \text{otherwise}.
    \end{cases}
\end{displaymath}

\noindent Analogously, $u(t+l-m\mid t)$ represents known, planned input values if the time index is in the future, and past measured inputs otherwise.

%==============================================================================================================================================================
\paragraph{Data Generation and System Excitation}
To produce the dataset $\mathcal{D}$, we excited the system by a combination of chirp and pseudo-random sequence (PRS) signals, generated in the \emph{dq} frame. These were used as additive perturbations on the reference of the current control loop, with amplitudes limited to $5\%$ of the base current. The signals were designed to span the frequency range $[0.1, 15]$ Hz, a bandwidth that encompasses the system's dominant electromechanical modes and the PLL dynamics \cite{hatziargyriou2020definition}, while remaining well below the Nyquist limit of the sampling frequency $f_s$. This choice ensures that the dynamics are persistently excited---thereby guaranteeing that $R$ in \eqref{eq:OP} is full row rank \cite{DePersisTesiCDC2019}---without introducing high-frequency noise that falls outside the spectral band of interest.

We wish to point out that such identification signals can have a non-negligible effect on the network's total harmonic distortion (THD). In practical applications, these can be scheduled so as not to cause continuous interference, thereby remaining compliant with the IEEE 519 standard; see \cite{blooming2006application, IEEE5191992}. 

By this experiment, we generated $29809$ input-output pairs, from which a training ARX dataset $\mathcal{D}$ of cardinality $N = 29798$ was obtained (allowing for a maximum model order $n_a=n_b=10$). The test dataset was obtained from an independent experiment, also using chirp-PRS signals, obtaining $2011$ input-output pairs to generate $N_t = 2000$ ARX elements. An additional validation set of $N_v = 10396$ elements was obtained from $10407$ input-output pairs by exciting the IBR's current reference loop with steps of random amplitude. These signals were chosen to mimic abrupt disturbances---such as sudden load variations or power imbalances---which present time-domain features different from the training data.

%==============================================================================================================================================================
\paragraph{Data Preprocessing and Local Validity Regions}
To ensure the numerical stability of the algorithm and the statistical validity of the correlation measures, the raw dataset $\mathcal{D}$ underwent a rigorous preprocessing pipeline. First, input and output data were normalized using z-score standardization to avoid scaling issues between variables with heterogeneous physical units. 

Subsequently, we addressed numerical scalability and stochastic non-stationarity. The dimension of the KKT system \eqref{eq:KKT_system_v}  associated with the complete dataset $\mathcal{D}$ would preclude its use for recursive prediction in real-time contexts (recall that matrix $\tilde{R}$ is dense). Furthermore, power system dynamics can exhibit distinct local behaviors depending on the operating point. To solve these issues, we partitioned the training dataset into $K=119$ regions, denoted as $\mathcal{D}_j$, $j=1, \dots, K$, using a balanced clustering strategy based on $k$-means over the regressor space. This yields homogeneous clusters of $N_{j} \approx 250$ points each; this is large enough to ensure statistical significance while still allowing efficient matrix operations.

\begin{figure}[tb]
    \centering
    \subfigure{
        \includegraphics[width=1\columnwidth,trim=0mm 0.6mm 0.2mm 3mm,clip]{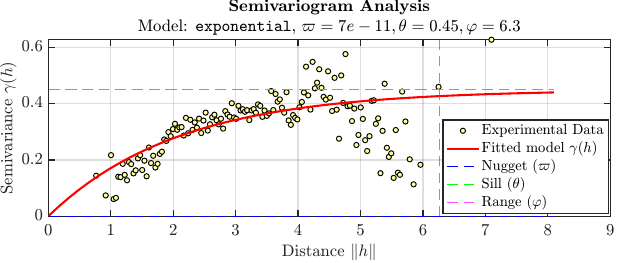}
    }

    \vspace{0mm}

    \subfigure{
        \includegraphics[width=1\columnwidth,trim=0mm 0.6mm 0.2mm 3mm,clip]{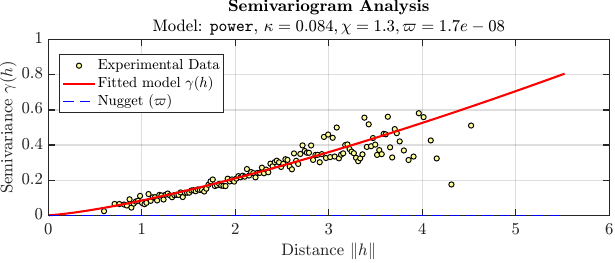}
    }

    \caption{Experimental semivariogram estimates (yellow markers) obtained by averaging pairwise semivariances within 200 distance bins, and the fitted theoretical models for clusters $j=10$ (top) and $j=114$ (bottom).}
    \label{fig:variogram_model}
\end{figure}

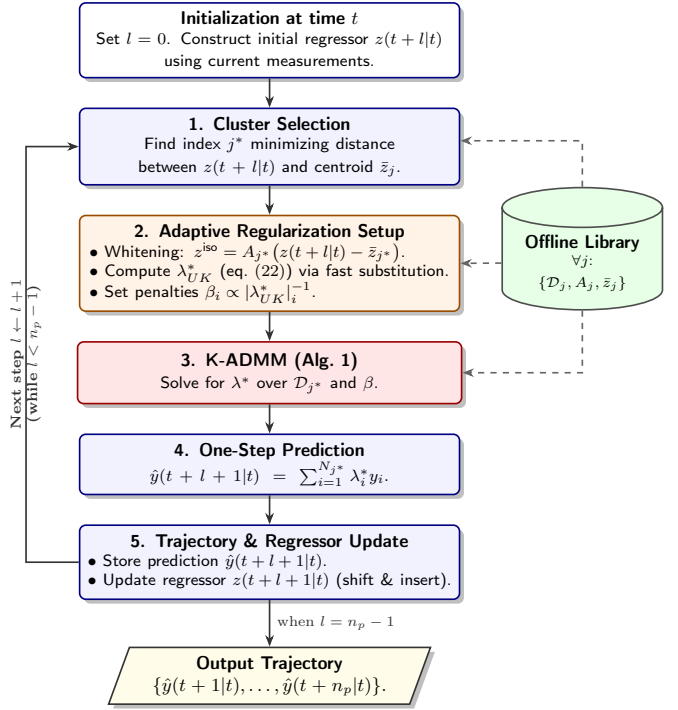
\begin{figure}[tb]
\centering
\resizebox{\columnwidth}{!}{
\begin{tikzpicture}[
    font=\sffamily\small,
    node distance=0.5cm and 0.5cm,
    >={Stealth[length=2mm]},
    process/.style={
        rectangle, 
        draw=blue!50!black, 
        thick, 
        fill=blue!5, 
        align=center, 
        text width=6.2cm, 
        minimum height=0.8cm, 
        rounded corners=3pt,
        inner sep=4pt,
        drop shadow
    },
    logic/.style={
        rectangle, 
        draw=orange!60!black, 
        thick, 
        fill=orange!10, 
        align=center, 
        text width=6.2cm, 
        rounded corners=3pt,
        inner sep=4pt,
        drop shadow
    },
    core/.style={
        rectangle, 
        draw=red!60!black, 
        thick, 
        fill=red!10, 
        align=center, 
        text width=6.2cm, 
        rounded corners=3pt, 
        inner sep=5pt,
        drop shadow
    },
    data/.style={
        cylinder, 
        shape border rotate=90, 
        aspect=0.25, 
        draw=black!70, 
        thick, 
        fill=green!10, 
        align=center, 
        text width=2.5cm, 
        minimum height=1cm
    },
    result/.style={
        trapezium, 
        trapezium left angle=70, 
        trapezium right angle=110, 
        draw=black!80, 
        thick, 
        fill=yellow!10, 
        align=center, 
        text width=4.5cm, 
        inner sep=4pt,
        drop shadow
    },
    line/.style={draw, thick, ->, color=black!80},
    dashed_line/.style={draw, thick, dashed, ->, color=black!60}
]
    \usetikzlibrary{shadows, positioning, fit, backgrounds, shapes.geometric}

    \node [process, fill=white] (init) {
        \textbf{Initialization at time $t$} \\
        \footnotesize 
        Set $l=0$. Construct initial regressor $z(t+l|t)$ \\
        using current measurements.
    };

    \node [process, below=0.5cm of init] (selector) {
        \textbf{1. Cluster Selection} \\
        \footnotesize 
        Find index $j^*$ minimizing distance \\
        between $z(t+l|t)$ and centroid $\bar{z}_j$.
    };

    \node [data, right=0.7cm of selector, anchor=west, yshift=-2.003cm] (db) {
        \textbf{Offline Library} \\ 
        \scriptsize 
        $\forall j$: \\
        $\{ \mathcal{D}_j, A_j, \bar{z}_j \}$
    };

    \node [logic, below=0.5cm of selector] (prior) {
        \textbf{2. Adaptive Regularization Setup} \\
        \footnotesize
        \begin{tabular}{@{\hspace{0pt}}l@{}}
         $\bullet$ Whitening: $z^\text{iso} = A_{j^*} \big(z(t+l|t) - \bar{z}_{j^*}\big)$. \\
         $\bullet$ Compute $\lambda^*_{UK}$ (eq.~\eqref{eq:kkt_system_local}) via fast substitution. \\
         $\bullet$ Set penalties $\beta_i \propto |\lambda^*_{UK}|_i^{-1}$.
        \end{tabular}
    };

    \node [core, below=0.5cm of prior] (kadmm) {
        \textbf{3. K-ADMM (Alg.~\ref{alg:ADMM_new})} \\
        \footnotesize 
        Solve for $\lambda^*$ over $\mathcal{D}_{j^*}$ and $\beta$.
    };

    \node [process, below=0.5cm of kadmm] (est) {
        \textbf{4. One-Step Prediction} \\
        \footnotesize 
        $\hat{y}(t+l+1|t) = \sum_{i=1}^{N_{j^*}} \lambda^*_i y_i.$
    };

    \node [process, below=0.5cm of est] (update) {
        \textbf{5. Trajectory \& Regressor Update} \\
        \footnotesize
        \begin{tabular}{@{\hspace{0pt}}l@{}}
         $\bullet$ Store prediction $\hat{y}(t+l+1|t)$. \\
         $\bullet$ Update regressor $z(t+l+1|t)$ (shift \& insert).
        \end{tabular}
    };

    \node [result, below=0.8cm of update] (final) {
        \textbf{Output Trajectory} \\
        $\{ \hat{y}(t+1|t), \dots, \hat{y}(t+n_p|t) \}.$
    };

    \draw [line] (init) -- (selector);
    \draw [line] (selector) -- (prior);
    \draw [line] (prior) -- (kadmm);
    \draw [line] (kadmm) -- (est);
    \draw [line] (est) -- (update);

    \draw [dashed_line] (db.north) |- ([yshift=0.1cm]selector.east);
    \draw [dashed_line] (db.west) -- (prior.east);
    \draw [dashed_line] (db.south) |- (kadmm.east);

    \draw [line] (update.west) -- ++(-0.9,0) |- 
        node [pos=0.35, left, font=\bfseries\scriptsize, rotate=90, align=center] 
        {Next step $l \leftarrow l+1$ \\ (while $l < n_p - 1$)} 
        (selector.west);

    \draw [line] (update.south) -- node [right, font=\scriptsize] {when $l = n_p - 1$} (final.north);

\end{tikzpicture}
}
\caption{Flowchart of the online recursive prediction procedure. At time step $t$, the regressor $z(t|t)$ is initialized. For each $l=0, \dots, n_p-1$, $\mathbf{1.}$ select the nearest data cluster $\mathcal{D}_{j^*}$, $\mathbf{2.}$ apply the pre-computed whitening transformation and compute the adaptive $\ell_1$ penalties using the LU factors, $\mathbf{3.}$ run the K-ADMM solver. Finally, the obtained prediction is incorporated into the regressor used in the subsequent iteration.}
\label{fig:online_flowchart}
\end{figure}

Within each data cluster $j$, we applied a geometric anisotropy correction. Specifically, we performed a whitening transformation ${z}^{\mathrm{iso}}_{i} = {A_j}(z_i - \bar{z}_j)$, for all sample locations $z_i$ considered in $\mathcal{D}_j$, where $\bar{z}_j$ is their average (the centroid regressor of the $j$-th cluster), and $A_j\in\mathbb{R}^{n\times n}$ is derived from principal component analysis~\cite[Sec.~2.5.2]{chiles2012geostatistics}.

%==============================================================================================================================================================
\paragraph{Variogram Modeling and Spectral Weighting} \label{sec:variogram_model_cs}
For each cluster, we first isolated the residuals $\delta(z^{\mathrm{iso}}_i)$ by subtracting the local trend $\mu_j^{\mathrm{iso}}$ from the observations. For this, we assumed the linear structure \eqref{eq:trend_model} and computed the corresponding parameters by ordinary least squares regression. We then fitted a semivariogram model to the residuals: for each cluster, we based our final choice on the coefficient of determination $\mathrm{R}^2$. In this case, either the \emph{exponential} or the \emph{power} models provided the best fit (for a total of 13 and 106 clusters, respectively).
The exponential variogram takes the form (see Fig.~\ref{fig:variogram_model})
\begin{equation}
    \gamma(h) = (\theta - \varpi) \left( 1 - \exp\left( - \frac{3\|h\|}{\varphi} \right) \right) + \varpi,
    \label{eq:exp_variogram}
\end{equation}
where $\theta$ represents the sill (total variance), $\varphi$ is the range parameter, and $\varpi$ denotes the nugget effect.\footnote{While the latter term breaks the theoretical definition \eqref{eq:th_variogram}, whereby $\gamma(0) = 0$, its role is to accommodate discontinuities due to measurement noise often observed near the origin in empirical semivariograms.}
The exponential model imposes a finite sill and range, and its linear behavior near the origin enables the modeling of continuous but non-differentiable processes, appropriate for capturing abrupt changes in the frequency derivative (ROCOF)~\cite[Sec.~2.3]{chiles2012geostatistics}. The power variogram reads as
\begin{equation}
    \gamma(h) = \kappa \|h\|^{\chi} + \varpi,
    \label{eq:power_variogram}
\end{equation}
where $\kappa$ determines the scale of the semivariance growth, $\chi$ its curvature, and $\varpi$ the nugget effect. Unlike the exponential model, the power variogram does not impose a finite sill or range, allowing it to fit data clusters whose empirical semivariogram keeps increasing over the observed distance range.
The suitability of these choices is empirically demonstrated in Fig.~\ref{fig:variogram_model}, which highlights the difference between second-order and intrinsically stationary data. To fit \eqref{eq:exp_variogram} and \eqref{eq:power_variogram} to the dataset, we developed a routine based on Python's \texttt{PyKrige} library \cite{pykrige_v173_2025}.
Finally, by evaluating the chosen variogram model at all coordinates $z_i^{\mathrm{iso}}$ in $\mathcal{D}_j$, we computed the corresponding local matrices $\Gamma_{\mathcal{D}_j}$, for $j=1,\dots,K$.

\begin{figure*}[t]
    \centering
    \subfigure{
        \includegraphics[width=0.48\textwidth,trim=4.2mm 2.0mm 11.5mm 3.7mm,clip]{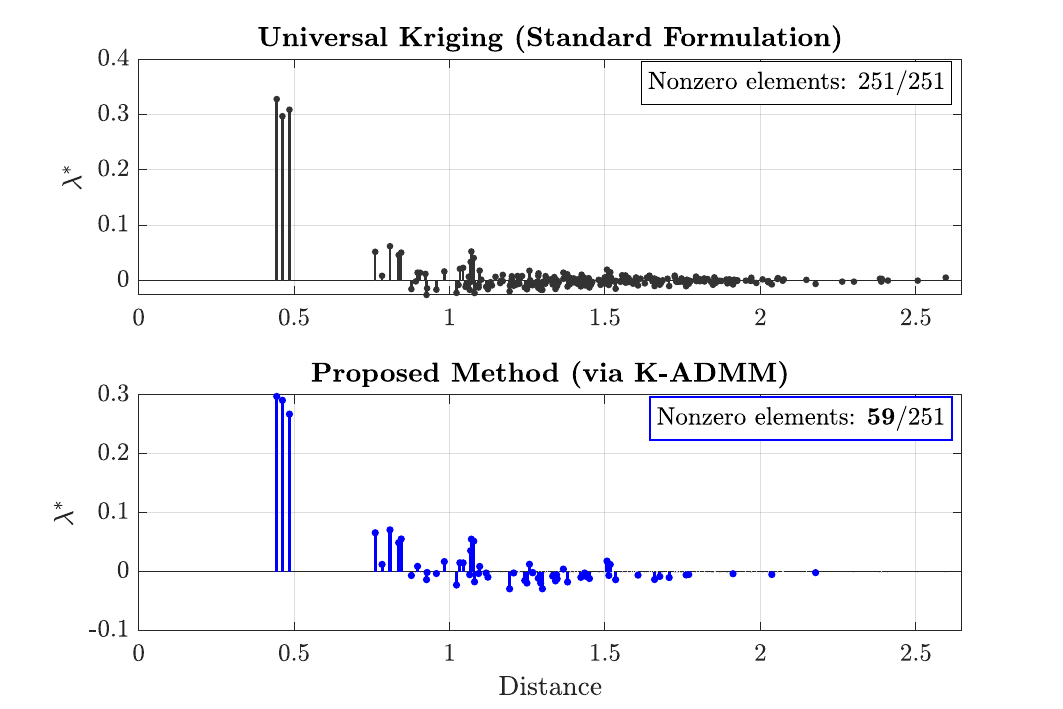}
    }
    \subfigure{
        \includegraphics[width=0.48\textwidth,trim=4.2mm 2.0mm 11.5mm 3.7mm,clip]{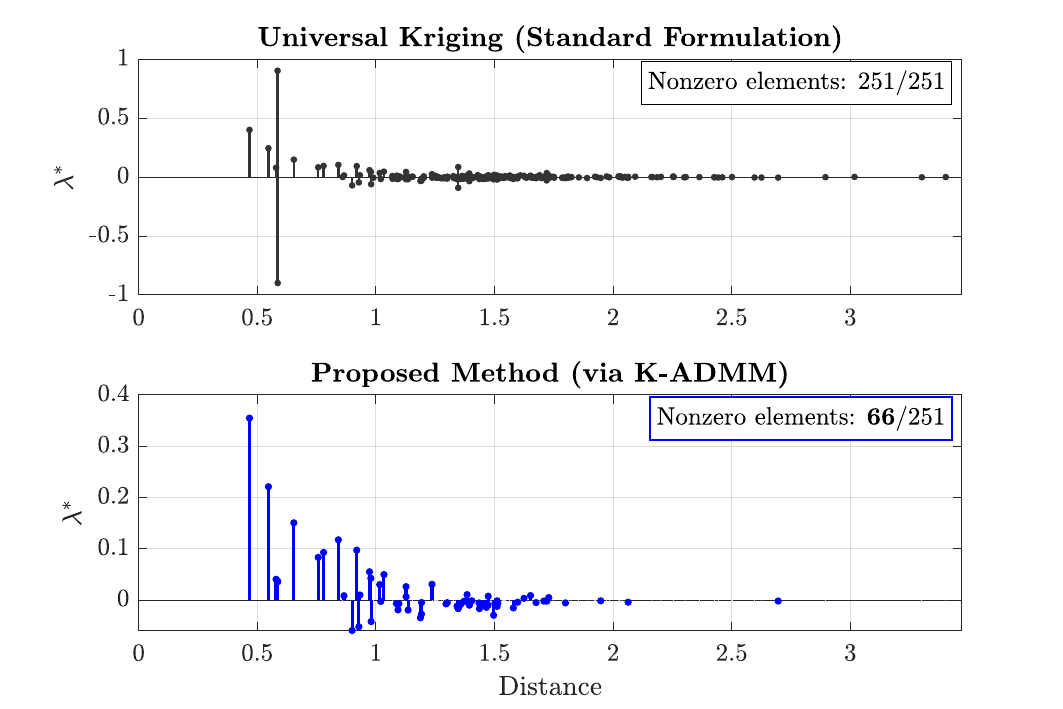}
    }
    \caption{Impact of the regularization $\ell_1$ on the $\lambda^*$ weights for two random query points (steps of trajectory). The top plots display the dense solutions obtained by the standard UK formulation, where the screening effect results in nonzero weights being assigned to the entire cluster. The bottom plots demonstrate how the proposed K-ADMM algorithm i) enforces sparsity, selecting a minimal subset of approximately $50$ highly informative neighbors ($\sim 75\%$ sparsity) while suppressing spurious correlations; and ii) produces lower prediction weight magnitudes.}
    \label{fig:esparsidad_UK_vs_UK_l1}
\end{figure*}

For each new query point, the sparsity-governing parameters $\beta$ were determined via the \textit{adaptive lasso} strategy \cite{matsui2025sparse}, where we used the standard (non-regularized) UK weights, $\lambda^*_\text{UK}$, as a prior. Specifically, we defined the individual penalties as $\beta_i = \varepsilon |\lambda^*_{\text{UK}}|_i^{-1}$, where $\varepsilon= 10^{-5}$ was obtained by cross-validation. Note that $\lambda^*_\text{UK}$ was obtained by solving the KKT system associated with \eqref{eq:OP_standard_QP}:
\begin{equation}\label{eq:kkt_system_local}
    \begin{bmatrix}
    -\Gamma_\mathcal{D} & R^\top \\
    R & 0
    \end{bmatrix}
    \begin{bmatrix}
    \lambda^*_\text{UK} \\ \varrho^*
    \end{bmatrix}
    =
    \begin{bmatrix}
    -\Gamma_0 \\ r_0
    \end{bmatrix},
\end{equation}
where $\varrho^*$ is the vector of Lagrange multipliers associated with the constraints. Since the matrix on the left-hand side depends solely on the training data within each validity region, its LU factorization can be computed offline. Consequently, the retrieval of the prior $\lambda^*_\text{UK}$ for any new query point reduces to a fast matrix-vector multiplication between the stored inverse factors and the dynamic vector $[-\Gamma_0^\top, r_0^\top]^\top$, obtained via the precomputed variogram.

The execution of this recursive strategy is depicted in Fig.~\ref{fig:online_flowchart}, which illustrates the complete sequence from regressor initialization $z(t \mid t)$ to generation of the full output trajectory.
We note that, once the regressor dimensions are fixed and all necessary matrices and parameters are precomputed offline, the computational cost of the online prediction loop scales linearly with $n_p$. (This has been empirically verified; however, the results are omitted for space reasons.)

%==============================================================================================================================================================
\paragraph{Benchmarking and model tuning}
To evaluate the predictor’s accuracy, the error of each trajectory was quantified using a discrete trapezoidal approximation of the relative error. Let $y_v(t+l)$ denote the true (test/validation) frequency and $\hat{y}(t+l \mid t)$ its predicted value at step $l$ within the $n_p$-step prediction horizon. The prediction error $\zeta(t)$ is defined as:
\begin{align} \label{eq:prediction_error}
\zeta(t)
  = & \frac{T_s}{2T} \sum_{l=1}^{n_p} \frac{\bigl|y_v(t+l)-\hat{y}(t+l\mid t)\bigr|}{\bigl|y_v(t+l)\bigr|} \nonumber\\
  & + \frac{T_s}{2T} \sum_{l=1}^{n_p} \frac{\bigl|y_v(t+l-1)-\hat{y}(t+l-1\mid t)\bigr|}{\bigl|y_v(t+l-1)\bigr|},
\end{align}
where $T_s = 1/f_s$ is the data sampling period, and $T = n_p T_s$ the total duration of the prediction horizon.

\begin{figure}[tb]
    \centering 
    \includegraphics[width=1\columnwidth, trim= 0mm 0.3mm 0.4mm 0mm,clip]{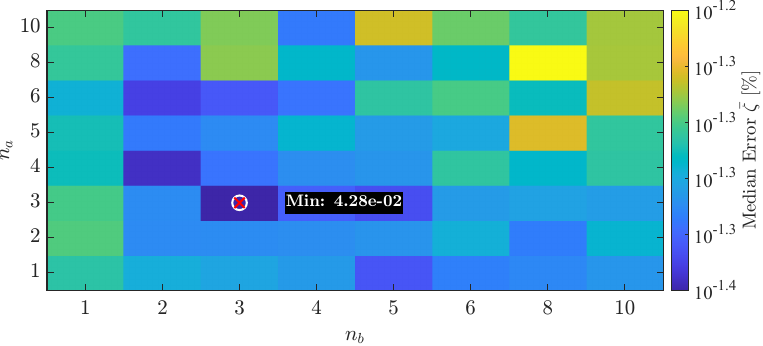}
	\caption{Heatmap of the total median error $\bar{\zeta}$ (in \%) over the test dataset, as a function of the autoregressive orders $n_a$ and $n_b$: the chosen configuration ($n_a=3, n_b=3$) minimizes the prediction error.}
	\label{fig:heat_map}
\end{figure}

\begin{figure*}[t]
    \centering
    \subfigure{
        \includegraphics[width=1\textwidth, trim=0mm 4.5mm 0.2mm 0mm,clip]{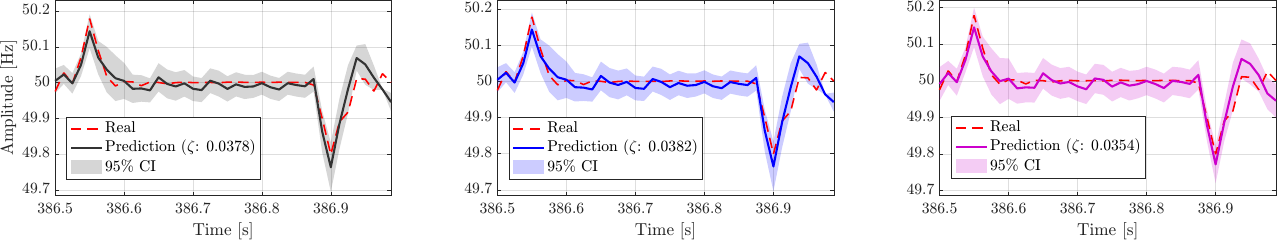}
    }\\
    \subfigure{
        \includegraphics[width=1\textwidth, trim=0mm 0.3mm 0.2mm -1mm,clip]{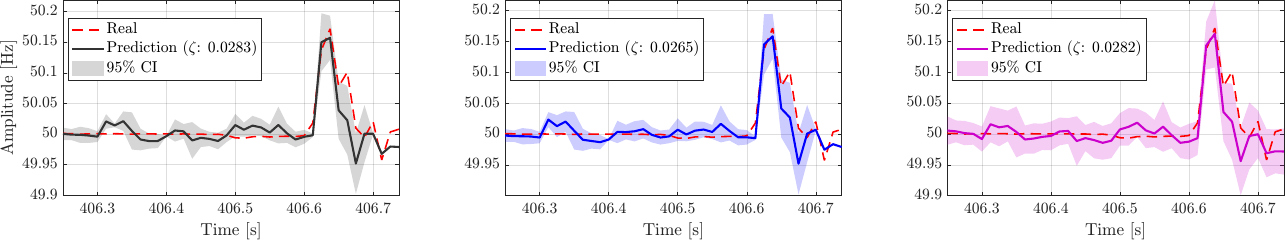}
    }
    \caption{Qualitative comparison of recursive frequency prediction over a $0.5$ s horizon across two distinct dynamic scenarios (rows). The columns contrast the performance of: (left) standard UK, (middle) the proposed {K-ADMM}, and (right) GP regression. The red dashed lines represent the ground truth, solid lines show the point predictions, and the shaded areas indicate  $95\%$ confidence intervals. Note that K-ADMM achieves comparable accuracy and uncertainty quantification while using a significantly sparser regressor.}
    \label{fig:matriz_predicciones_UK_vs_ADMM_vs_GPR}
\end{figure*}

We obtained appropriate values for $n_a$ and $n_b$ by cross-validation on the test dataset. The heatmap in Fig.~\ref{fig:heat_map} shows how the combination $n_a = 3$ and $ n_b = 3$ yielded the best prediction accuracy: the resulting regressor is of dimension $12$, taking into account the future input sequence introduced in each recursive regressor. Finally, the parameter $\rho$ was also adjusted to a value of $0.1$ using the test dataset in order to achieve the best convergence rate.

\begin{figure}[tb]
    \centering
    \includegraphics[width=1\linewidth, trim=0mm 0.1mm 0mm 0mm,clip]{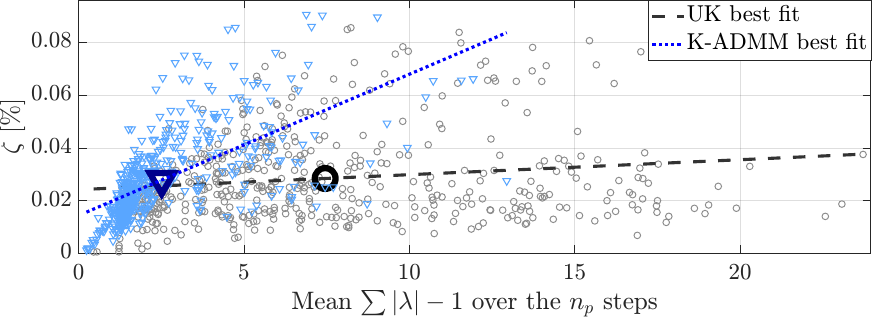}
    \caption{Impact of regularization on weight quality and prediction accuracy. The metric $\sum |\lambda_i| - 1$ quantifies the magnitude of negative weights. The analysis shows that negative weights correlate with higher errors (as visible through the best-fit slopes). K-ADMM avoids this by promoting interpolation (reflected by lower values of $\sum |\lambda_i| - 1$) while maintaining comparable prediction accuracy (as shown by the relative locations of the centroids, denoted by large markers).}
    \label{fig:sparsity_vs_error}
\end{figure}

\begin{figure}[tb]
    \centering
    \includegraphics[width=0.9\columnwidth, trim=0mm 0mm 0.2mm 0mm,clip]{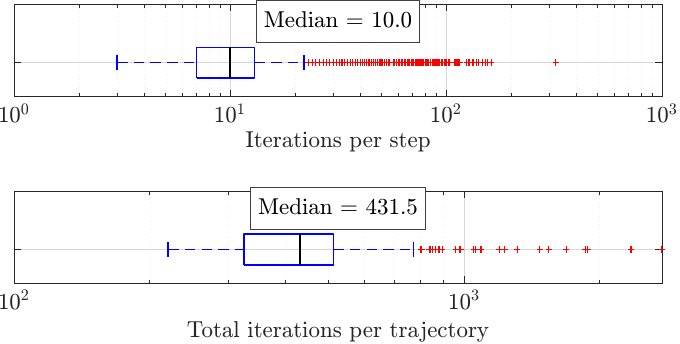}
    \caption{Convergence of the K-ADMM algorithm across the validation dataset. The box-plots illustrate the distribution of iterations required to satisfy the stopping criterion ($\epsilon^{\text{pri}} = \epsilon^{\text{dual}} = 10^{-5}$). \textit{Top plot}: Iterations per individual prediction step, showing a median of only $10$ iterations. \textit{Bottom plot}: Iterations required to compute the complete $0.5$-second trajectory (median: $431.5$ iterations).}
    \label{fig:convergence_iterations}
\end{figure}

%==============================================================================================================================================================
\paragraph{Results}
We evaluated the computational performance on a machine equipped with an Apple Silicon M3 Pro processor and 18 GB of RAM, where the whole procedure---detailed in Fig.~\ref{fig:online_flowchart} (including Alg.~\ref{alg:ADMM_new})---was implemented through standard MATLAB scripts (2024a release).

To demonstrate the effect of adaptive regularization, Fig.~\ref{fig:esparsidad_UK_vs_UK_l1} compares the distribution of the optimal weights $\lambda^*$ obtained by standard UK (top row) against K-ADMM (bottom) for two distinct query points (left and right panels). While the standard formulation assigns nonzero weights to all points belonging to the local data cluster, K-ADMM effectively suppresses irrelevant correlations. As observed, a sparse subset of approximately $50$ nonzero neighbors is selected (achieving $\sim 75\%$ sparsity), enhancing the model's interpretability and robustness.
To achieve exact sparsity, all elements of $\lambda^*$ with magnitude below $10^{-4}$ were truncated to zero (cf.~Sec.~\ref{sec:ADMM_out}); this threshold was chosen to be more than an order of magnitude smaller than $1/N_j \approx 4 \cdot 10^{-3}$, ensuring that the discarded information was statistically negligible.

To further validate
the proposed approach, Fig.~\ref{fig:sparsity_vs_error} correlates the prediction error \eqref{eq:prediction_error} with the interpolation metric $\sum |\lambda_i| - 1$ (where 0 implies pure interpolation). In particular, the horizontal coordinates in Fig.~\ref{fig:sparsity_vs_error} correspond to the average metric over the $n_p$ steps which compose each prediction. 
The positive slope of the dashed lines (linear best fits of the points) indicates that  higher magnitudes of negative weights correlate with higher prediction errors. Indeed, standard UK (black circles) spreads widely to the right, revealing its reliance on large negative weights, whereas K-ADMM (blue triangles) concentrates closer to zero, reflecting the promotion of predictions obtained through convex combinations of data points (interpolation). 
Also, the relative position of the centroids---with the K-ADMM one shifted significantly to the left---indicates that K-ADMM affords a median error comparable to or lower than UK. From these tests, we can infer that  regularization removes artifacts and improves  interpretability without sacrificing accuracy.

We then compared the proposed approach with GP regression. For the latter, we also considered the model \eqref{eq:trend_model} to capture the trend, and used the exponential kernel 
   $\sigma_f^2 \exp\left( - \| h \|/\sigma_l \right)$,
where $\sigma_f^2$ is the variance of the residual and $\sigma_l$ is the characteristic length scale \cite{Rasmussen2006gaussian}.
A kernel of this form was fitted to each of the training datasets $\{\mathcal{D}_j\}_{j=1}^K$ (the same used for kriging), via MATLAB's \texttt{fitrgp} function, using the sparse fully-independent conditional approximation method, restricted to an active set size equal to $100$ to limit computation time.

Fig.~\ref{fig:matriz_predicciones_UK_vs_ADMM_vs_GPR} provides a qualitative comparison of the predicted trajectories in two dynamic transient scenarios (rows). The columns contrast the performance of non-regularized UK (left), K-ADMM (middle), and GP regression (right). All three methods successfully capture the non-linear transient dynamics and provide similar 95\% confidence intervals (shaded region). Note that these intervals are intrinsic only to the GP formulation: for UK and regularized UK, nominal $95$\% prediction intervals are constructed by assuming a Gaussian random process \cite[Sec.~3.4]{cressie1993statistics}. All three methods attained comparable prediction errors across the validation dataset. In particular, K-ADMM yielded a median error across the validation dataset of $\bar{\zeta}=0.0238\%$, which is similar to those obtained by non-regularized UK ($\bar{\zeta}=0.0244\%$) and GP regression ($\bar{\zeta}=0.0238\%$), confirming that sparsity does not compromise accuracy.
It should be noted that K-ADMM attains this predictive fidelity using only a fraction of the information (see Fig.~\ref{fig:esparsidad_UK_vs_UK_l1}).

\begin{table}[tb]
    \centering
    \small
    \begin{tabular*}{0.9\linewidth}{l @{\extracolsep{\fill}} *{3}{r}}
    \toprule
    \multirow{2}{*}{\textbf{Method}} & \multicolumn{3}{c}{\textbf{Computation time} [ms]} \\
    & \multicolumn{1}{c}{max} & \multicolumn{1}{c}{median} & \multicolumn{1}{c}{min} \\
    \midrule
    \textbf{K-ADMM}   & 231.29 & 39.24 & 15.48 \\
    \textbf{quadprog} & 18441.7  & 8130.97  & 1794.15 \\
    \bottomrule
    \end{tabular*}
    \caption{Computation time on the validation dataset. The times for K-ADMM include the entire recursive procedure illustrated in Fig.~\ref{fig:online_flowchart}. A standard QP reformulation of \eqref{eq:OP_regularized}, using auxiliary variables, is solved through MATLAB's \texttt{quadprog}.}
    \label{tab:results_summary}
\end{table}

Finally, as regards numerical efficiency, we compared our method against the solution  of the standard QP reformulation of \eqref{eq:OP_regularized} (see, e.g., \cite[Sec.~6.1]{boyd2004convex}) using MATLAB's \texttt{quadprog}. The latter required a median time of $8.131$~s to compute the entire prediction, as opposed to the $39.24$~ms achieved by using K-ADMM. Additional details are available in Table~\ref{tab:results_summary}: the reported times include the entire recursive procedure schematized in Fig.~\ref{fig:online_flowchart}. This involved a median computational cost of $10$ iterations until convergence for Alg.~\ref{alg:ADMM_new}, for each of the $n_p = 40$ steps composing the full trajectory (see Fig.~\ref{fig:convergence_iterations}).

%============================================================================================================================================================================================================================================================================================================================
\section{Conclusion}
This paper introduces a data-driven kriging framework for short-term frequency forecasting in low-inertia power systems. 
We propose an efficient solution algorithm (K-ADMM) which mitigates numerical ill-conditioning while enforcing sparsity in the predictor, by incorporating adaptive $\ell_{1}$-regularization into the kriging problem.

The approach was validated on a simulated case study modeling an inverter-based device (IBR) connected to a weak grid. Results show that the proposed predictor accurately captures nonlinear transients and abrupt frequency variations (ROCOF) over a $0.5$-second horizon, achieving a median prediction error of $0.024\%$. Performance is comparable to standard UK and GP regression, with substantially lower model complexity ($\sim 75\%$ sparsity). Numerical tests show that K-ADMM attains median computation times of $39$~ms per predicted trajectory, making it a viable tool for FFR applications.

%============================================================================================================================================================================================================================================================================================================================
\section*{Acknowledgements}
This work was supported in part by 
grant PID2022-142946NA-I00 funded by MICIU/AEI/ 10.13039/501100011033 and by ERDF/EU, 
and in part by 
the R\&D project Learning-based robust control of smart infrastructures with certified reliability (ref.~2024/00000800), co-funded by EU--Spanish Ministry of Finance--European Funds--Junta de Andaluc\'ia--Consejer\'ia de Universidad, Investigaci\'on e Innovaci\'on. F.~Fele also gratefully acknowledges support from grant RYC2021-033960-I funded by MICIU/AEI/10.13039/501100011033 and European Union NextGenerationEU/PRTR.

%============================================================================================================================================================================================================================================================================================================================
\bibliographystyle{elsarticle-num}
{\small
\bibliography{bib.bib}}

%============================================================================================================================================================================================================================================================================================================================
\appendix
%==============================================================================================================================================================
\section{Proof of Proposition \ref{prop:optimal:z}}
\label{Appendix:Proof:Optimal:z}
Let $\Phi(x)\coloneqq wx^2 + \beta |x| - cx$ denote the objective function in \eqref{eq:family_problem}, and notice that it is a strictly convex function in $x$ (since $w>0$ and $\beta\geq 0$) which admits a unique minimizer $x^*$. From the optimality of $x^*$ we have $$ 0 =  \Phi(0) \geq \Phi(x^*) = w(x^*)^2 + \beta|x^*| - c x^* \geq - c x^*.$$
Therefore, $c x^* {\geq} 0$, which implies $c x^* = |c| |x^*|$. 
Thus, $$ \Phi(x) = w x^2 + (\beta-|c|)|x|, $$
and $$ |x^*|= \argmin\limits_{s\geq 0} w s^2 + (\beta-|c|)s.$$
By inspection, we infer that $|x^*|=0$ if $\beta-|c|\geq 0$. Otherwise, $|x^*| = \frac{|c|-\beta}{2w}$. Therefore, $|x^*|=\frac{1}{2w}\max\{0, |c|-\beta\}$. This, along with $c x^* {\geq} 0$, yields $x^* = \frac{1}{2w}\mathrm{sign}(c)\max\{0, |c| - \beta\}$, concluding the proof. \hfill \qed

We note that this solution is also equivalent to the soft-thresholding operator, i.e., the $\ell_1$-norm proximal operator; see, e.g., \cite[Example 6.8]{beck2017first}.

%============================================================================================================================================================================================================================================================================================================================
\end{document}